\documentclass[useAMS,usenatbib]{mn2e}

\usepackage{txfonts}
\usepackage{graphicx}
\usepackage{epstopdf}
\usepackage{natbib}
\newcommand{\teffm}{$T_{\rm eff}$}
\newcommand{\teff}{\rm T_{eff}}
\newcommand{\msunm}{$\rm M_\odot$}
\newcommand{\mstarm}{$\rm M_*$}
\newcommand{\logg}{$\log \rm g$}
\newcommand{\Menvm}{$\rm log(M_{env})$}
\newcommand{\Menv}{\rm log(M_{env})}
\newcommand{\Mhem}{$\rm log(M_{He})$}
\newcommand{\qfmm}{$\rm q_{fm}$}
\newcommand{\Xom}{$\rm X_{o}$}
\newcommand{\sigrmsm}{$\rm \sigma_{RMS}$}

\title[Asteroseismic constraints on diffusion in WD envelopes] {Asteroseismic constraints on diffusion in white dwarf envelopes}
\author[A. Bischoff-Kim and T. S. Metcalfe]{A. Bischoff-Kim$^{1}$\thanks{E-mail:agnes.kim@gcsu.edu} 
and T. S. Metcalfe$^{2}$\\
$^{1}$Georgia College \& State University, Department of Chemistry, Physics and Astronomy, CBX 082, Milledgeville, GA 31061-0490, USA \\
$^{2}$High Altitude Observatory, National Center for Atmospheric Research, PO Box 3000, Boulder, CO 80307-3000, USA}

\begin{document}

\maketitle

\label{firstpage}

\begin{abstract}
The asteroseismic analysis of white dwarfs allows us to peer below their photospheres and determine their internal structure.  At $\sim 28,000$ K EC20058-5234 is the hottest known pulsating helium atmosphere white dwarf. As such, it constitutes an important link in the evolution of white dwarfs down the cooling track. It is also astrophysically interesting because it is at a temperature where white dwarfs are expected to cool mainly through the emission of plasmon neutrinos. In the present work, we perform an asteroseismic analysis of EC20058-5234 and place the results in the context of stellar evolution and time dependent diffusion calculations. We use  a parallel genetic algorithm complemented with targeted grid searches to find the models that fit the observed periods best. Comparing our results with similar modeling of EC20058-5234's cooler cousin CBS114, we find a helium envelope thickness consistent with time dependent diffusion calculations and obtain a precise mode identification for EC20058-5234.
\end{abstract}


\section{Astrophysical context}

White dwarfs are the end product of the evolution of around 98\% of the stars. Buried
in their interiors are the records of physical processes that take place during
earlier stages in the life of the star. Nuclear reaction rates during the core 
helium burning phase set the core composition of white dwarfs, while the 
relative time spent burning hydrogen and helium during the AGB phase and 
massloss episodes determine the thickness of the helium layer 
\citep{Lawlor06,Althaus05}. 

%%Text removed%%%%%%%%%%
%Time dependent diffusion calculations 
%\citep{Dehner95,Althaus05} make specific predictions on the evolution of
%the shape of the helium composition profiles as the interior cools. Namely, we expect
%the thickness of the pure Helium layer surrounding the core to increase with time. This 
%means that cooler white dwarfs, being older, should have thicker Helium layers. The asteroseismology
%of pulsating white dwarfs allows us a way to determine their interior structure and to test 
%this theory.

%%Text added%%%%%%%%%%%%%%%%%%%%%%%%%%%%%%
Helium atmosphere white dwarfs (DBs) comprise roughly 20\% of the population of field white dwarfs,
with most of the remaining 80\% consisting of their hydrogen atmosphere (DA) cousins. The majority of white dwarfs in both
of these spectral classes are thought to arise from the evolution of isolated main-sequence stars with masses that are insufficient to ignite carbon fusion, which ultimately leave their hot carbon/oxygen cores to descend the white dwarf cooling
track. The bifurcation into two spectral classes is thought to occur during post-asymptotic-giant-branch (post-AGB) evolution when, in some cases, a very late thermal pulse burns off the residual hydrogen in the envelope, producing a nearly pure
helium atmosphere \citep{Iben83}. Such objects are then supposed to return to the white dwarf cooling track as PG 1159
stars, which are widely believed to be the precursors of most DB white dwarfs.

If we assume that there is an evolutionary connection between the PG 1159 stars and the cooler DB white dwarfs,
we can look to several independent groups who have used time-dependent diﬀusion calculations to follow the changes in
the interior structure of these objects as they cool \citep{Dehner95, Fontaine02, Bertolami06}. The hot PG 1159 stars, having recently emerged from the born-again phase, contain envelopes with a nearly uniform
mixture of helium (He), carbon (C), and oxygen (O) out to the photosphere \citep{Dreizler98, Herwig99}. As
they cool, the helium diffuses upward and gradually accumulates to form a chemically pure surface layer. Through the DBV instability strip, this process is still ongoing so that instead of a pure helium layer surrounding the carbon and oxygen core, one has a region where the carbon and helium are still mixed. This leads to a double-
layered structure, with the pure He surface layer overlying the remainder of the uniform He/C/O envelope, all above the
degenerate C/O core (see Fig \ref{f1}).

A key prediction of the diffusion models is that, for a given stellar mass, the pure He surface layer will steadily
grow thicker as the DB star cools. The only available observational tests of this prediction come from asteroseismology -- the study of the internal structure of stars through the interpretation of their pulsation periods. %%%%%%%%%%%%%%end text added%%%%%%%%%%%%%%%%%%%%
Helium atmosphere white dwarfs (DBs) are found to pulsate at effective temperatures ranging between 21,000 K and 28,000 K \citep{Beauchamp99,
Castanheira05}. There are currently 20 known pulsating DBs (DBVs) \citep{Nitta09, Kilkenny09}. To date two of them, GD358 
and CBS114, have been the object of detailed asteroseismic analyses. When it is well behaved, GD358 is the poster child for white dwarf asteroseismology, offering for analysis 11 dipole modes with consecutive radial overtones k ranging between 8 and 18 \citep{Bradley94}. However, an ambiguity between structure in the core and structure in the envelope has made it 
difficult to uniquely determine its core structure \citep{Montgomery03}. The  observed pulsation spectrum of CBS114 also contains 11 dipole modes, ranging from k=8 to 20. \citet{Metcalfe05a} performed a systematic asteroseismic analysis using a genetic algorithm to search for best fit models of this star.  The models used in that work included realistic carbon and oxygen core abundance profiles that were allowed to vary to find the best fit.

While the ambiguity in the interior structure of GD358 made it impossible to get a clear picture of its core chemical abundance profile, \citet{Metcalfe07} was able to compare asteroseismic analyses of GD358, CBS114 and EC20058-5234 that assumed pure carbon cores. The analysis showed a correlation between the thickness of the helium layer and the effective temperature of the models, with EC20058-5234 having the lowest helium layer mass. While the result was consistent with the predictions of time dependent diffusion calculations, there is no physical justification to assume pure carbon cores. It is not consistent with stellar evolution calculations and core structure influences asteroseismic fits. In the present study we perform a more physical asteroseismic fit of EC20058-5234 to compare with the CBS114 fits done by \citet{Metcalfe05a}, where non-zero core oxygen abundances were considered. 

EC20058-5234 was the target of a Whole Earth Telescope observing campaign  \citep[WET;][]{Nather89}
that revealed 8 or 9 stable, independent modes \citep{Sullivan08} . With that 
many modes, we can expect the asteroseismology of EC20058-5234 to yield useful 
constraints on its properties. We perform the first detailed asteroseismic analysis of EC20058-5234 to 
determine its stellar parameters and internal properties. In Sect. \ref{section2}, we summarize the clues we used from spectroscopy to guide our asteroseismic analysis. In Sect. \ref{section3}, we describe our models and the method we followed. We present our results in Sect. \ref{section4} and in Sect. \ref{section5}, discuss our results in the framework of stellar evolution. We conclude in Sect. \ref{section6}.

\section{Clues from observations}
\label{section2}

\subsection{Spectroscopy}

Assuming a pure helium atmosphere, \citet{Beauchamp99} determined an 
effective temperature for EC20058 of 28,400 $\pm$ 1,500~K and a \logg~of 7.86 $\pm$ 0.10. 
DB white dwarfs present a unique challenge to spectroscopists because of the 
lack of hydrogen lines in their spectra. Because no spectrum is free of noise, 
it is possible to add in the models trace amounts of hydrogen that do not result in 
detectable lines in the spectrum. Unfortunately, even trace amounts of hydrogen
can lead to significant differences in the inferred effective temperature and 
\logg. Introducing a maximum trace amount of hydrogen N(He)/N(H) = -3.5, 
\citeauthor{Beauchamp99} determined an effective temperature of 27,100 $\pm$ 
1,500~K and a \logg~of 7.80 $\pm$ 0.10. The spectroscopic ``box'' for EC20058-5234 
is therefore ($25,600~\rm K < \teff < 29,900~K, 7.70 <$ \logg~$< 7.96$). With our models, the \logg~ constraint translates to a mass constraint 0.46 \msunm $<$ \mstarm $<$ 0.60 \msunm. EC20058-5234 is a hot DBV and has a lower than average mass for a white dwarf.

\subsection{Pulsation spectrum}
\label{pspectrum}

EC20058-5234 was first discovered to pulsate by \citet{Koen95}. In a 1997 
Whole Earth Telescope run on this star, 
\citet{Sullivan08} found 11 fundamental modes, listed in Table \ref{pobs}. The two 
highest amplitude modes f6 (281.0s) and f8 (256.9s) are remarkably stable and 
attempts have been made to use them to measure a cooling rate for EC20058-5234 \citep{Dalessio10}.

\begin{table}
\caption{
\label{pobs}
Observed periods in EC20058-5234}
\centering
\begin{tabular}{|c|c|l|}
%\tableline
\hline
\rule[-0.2cm]{0mm}{0.6cm}
 Mode name         & Period [s]  & Notes    \\
\hline
\rule[-0.0cm]{0mm}{0.6cm}
f1  & 539.8 &                                \\
f2  & 525.4 &                                \\
f3  & 350.6 &                                \\
f4  & 333.5 &                                \\
f5  &(286.6)& m=+1 rotational split of f6    \\
f6  & 281.0 & High amplitude, stable mode, \\
     &            & photometrically identified as $\ell=1$  \\
f7  & 274.7 & Possible m=-1 rotational split of f6 \\
f8  & 256.9 & High amplitude, stable mode    \\
f16 &(207.6)& m=+1 rotational split of f9    \\
f9  & 204.6 &                                \\
f11 & 195.0 &                                \\
\hline
\end{tabular}
\end{table}

\citeauthor{Sullivan08} found two modes each split by 70 $\rm \mu$Hz (f6 and f9). If we 
assume these modes are $\ell$=1 modes (a reasonable assumption from asymptotic 
period spacing arguments and the relatively low \logg~found from spectroscopy), 
then the 70 $\rm \mu$Hz splitting is consistent with a rotation period of the 
star of 2 hours. The amplitude ratios of harmonics and combination peaks to parent modes supports the hypothesis that f6 is an $\ell=1$ mode \citep{Yeates06}. Since f9 has an identical frequency split likely due to rotation, it is reasonable to assume that it is an $\ell=1$ mode as well. \citeauthor{Sullivan08} also suggest that perhaps f7 is the third member of the (f5,f6,f7) rotationally split $\ell$=1 triplet, invoking the existence of a 3 kG magnetic field to account for the uneven frequency splitting. f7 could also be a mode of its own that happens to lie close to where the m=+1 member of the (f5,f6) multiplet would be if it were present. Preliminary work \citep{Bischoff-Kim08c} did not find that fits excluding the f7 mode (e.g. invoking a magnetic field) were 
significantly better than fits that included the f7 mode as an independent mode. In the 6 parameter asteroseismic fits presented in this work, we consider both alternatives. 

\section{The models}
\label{section3}

For this work, we used  a parallel genetic algorithm applied to white dwarf asteroseismology \citep{Metcalfe03a}, complemented with targeted grid searches. The genetic algorithm is an efficient way to search vast areas of parameter space to find local minima. Grid searches around these areas allow us to refine our results and produce a picture of parameter space around these local minima. To compute all our models, we used the White Dwarf Evolution Code (WDEC).

The WDEC evolves hot polytrope models from temperatures close to 100,000K down to the temperature of our choice. Models in the temperature range of interest for the present study are thermally relaxed solutions to the stellar structure equations. Each model we compute for our grids is the result of such an evolutionary sequence. 

\subsection{Input physics}

The WDEC is described in detail in \citet{Lamb75} and \citet{Wood90}. We used 
smoother core composition profiles and experimented with the more complex 
profiles that result from stellar evolution calculations \citep{Salaris97}. We updated the envelope equation of state tables from those calculated by 
\citet{Fontaine77} to those given by \citet{Saumon95}. We use OPAL opacities 
\citep{Iglesias96} and plasmon neutrino rates published by \citet{Itoh96}. The new envelope equation of state tables and plasmon neutrino rates are input physics we updated since the work of \citet{Metcalfe05a}.

\begin{figure}
  \resizebox{\hsize}{!}{\includegraphics{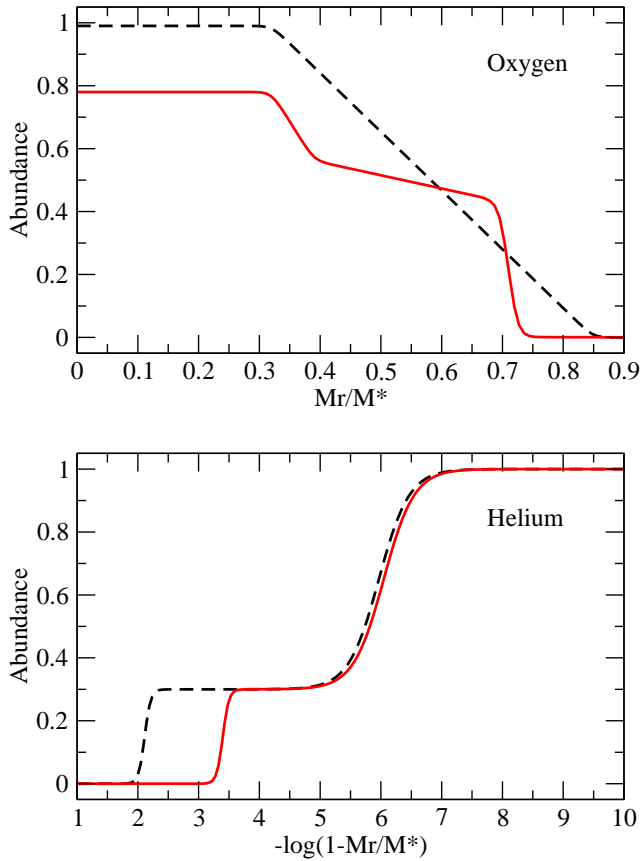}}
  \caption{Oxygen and helium abundance profiles for the Salaris like (solid lines) and ramp (dashed lines) best fit models.}
  \label{f1}
\end{figure}

DBV's are younger than their cooler cousins the DAV's. Time dependent diffusion calculations show that at 24,000 K, a typical temperature for a DBV, the carbon has not fully settled into the core of the star yet, and we expect double layered helium layers, as shown in Fig. \ref{f1} (the chemical profiles corresponding to our best fit model for EC20058-5234). Following  \citet{Metcalfe05a}, we adopted and parameterized this structure in our models. \Menvm~marks the location of the base of the helium layer and \Mhem~marks the location where the helium abundance rises to $1$. \Menvm~and \Mhem~are mass coordinates, defined as e.g. $\Menv = -\log(1 - M(r)/M_*)$, where M(r) is the mass enclosed in radius r and $\rm M_*$ is the stellar mass. We did not treat the helium abundance in the carbon/helium region as a free parameter, but adopted the values predicted from time dependent diffusion calculations in \citet{Dehner95}.

There are two parameters associated with the shapes of the oxygen (and 
carbon) core composition profiles: the central oxygen abundance ($\rm X_o$), and the edge of the homogeneous carbon and oxygen core (\qfmm).  We show an example of a basic oxygen abundance profile in Fig. \ref{f1}
along with a Salaris-like profile. We tried both the simple profiles and Salaris-like profiles for the carbon and oxygen abundances, varying the parameters \Xom and \qfmm.
 
\subsection{Global exploration of parameter space with the Genetic Algorithm}

To find a best fit model to the observed periods of EC20058-5234, we varied 6 parameters in the ranges listed in Table 2. The effective temperature and mass range were decided based on the spectroscopy.

\begin{table}
\caption{
Parameters varied in the fits }
\centering
\begin{tabular}{|c|c|l|}
\hline
 Parameter & Range  & Description    \\
\hline
\teffm        & 25000 - 30000 K                    & Effective temperature             \\
\mstarm     & 0.450 - 0.575 \msunm           & Stellar mass                             \\
\Menvm     & -2.00 to -4.00                        & Location of the base of           \\
                  &                                               & the helium layer                       \\
\Mhem       & -5.00 to -7.00                        & Location of the carbon/helium \\
                  &                                               & to pure helium transition zone \\
\qfmm        & 0.10 - 0.85 $\rm M_{*}$  & Location of the edge of           \\
                  &                                               & the homogeneous C/Ocore      \\
\Xom          & 0.0 - 1.0                                & Central Oxygen abundance     \\ 
\hline
\end{tabular}
\end{table}

The goal in white dwarf asteroseismology is to find the white dwarf model whose periods best agree with the observed periods. The genetic algorithm is able to explore the more promising regions of parameter space, and find the neighborhood of a global minimum. This supposes the existence of a global minimum. With 8 or 9 periods for 4 or 6 parameters, the problem is usually well constrained and the genetic algorithm successful. For EC20058-5234, the genetic algorithm was able to find a unique, global minimum. 

To calibrate the models used here against those used by \citet{Metcalfe05a}, we started with pure carbon core models ($\rm X_o=0$). For such fits, 4 parameters describe the models fully: the effective temperature, the mass, and the 2 envelope structure parameters \Menvm~and \Mhem. We know from stellar evolution that nuclear burning in the phases leading up to the white dwarf stage results in cores made up not only of carbon, but also of oxygen \citep[e.g.][]{Althaus10}. The C/O abundance ratio depends on the $C(\alpha,\gamma)O$ reaction cross-section. We also know \citep{Montgomery03} that asteroseismic fits are most sensitive to chemical transitions in the core. In light of these facts, more recent asteroseismic fits have considered varied core compositions \citep[e.g.][]{Metcalfe01,Bischoff-Kim09}. We performed such fits, varying the 2 core parameters $\rm X_o$ and \qfmm~in addition to the 4 parameters of the pure carbon models. We tried both basic core abundance profiles with an artificial, linear decrease in oxygen with radius ("ramp" profiles) and more physically realistic profiles, based on the work by \citet{Salaris97}.

As mentioned in Sect. \ref{pspectrum}, we also have two different hypotheses concerning the number of independent modes we have at our disposal to fit. If one believes that the 274.7~s mode is the result of an uneven rotational split of the 281~s mode, then we have only 8 independent modes. If one rejects that hypothesis, then there are 9 independent modes to fit. We tried both possibilities. For clarity and future reference, we summarize the different kinds of models considered in table 3.

\begin{table}
\label{modeltypes}
\caption{
A summary of the different models considered }
\centering
\begin{tabular}{|l|l|l|}
\hline
\rule[-0.2cm]{0mm}{0.6cm}
Name	& Description                                  & Parameters					        \\
\hline
Pure C	& Pure carbon core                         & \teffm, \mstarm, \Menvm, \Mhem		        \\
Ramp	& C and O core  & \teffm, \mstarm, \Menvm, \Mhem, \Xom, \qfmm     \\	
Salaris	& \citet{Salaris97}                          &                                                                                  \\
            & C and O core   &   \teffm, \mstarm, \Menvm, \Mhem, \Xom, \qfmm   \\  
\hline
\end{tabular}
\end{table}

\section{Results}
\label{section4}

\begin{table}
\label{results}
\caption{
Best fit parameters and goodness of fit. BIC (Bayes Information Criterion) is defined in the text. }
\centering
\begin{tabular}{|l|llllll|l|l|}
\hline
	& \multicolumn{8}{|c|}{8 periods}		\\
\hline
Model	& \multicolumn{6}{c|}{\teffm [K], \mstarm [\msunm], \Menvm, \Mhem, \Xom, \qfmm} & \sigrmsm &  BIC \\
\hline
Pure C 	& 29600 & 0.530 & -3.46 &  -6.26   &           &            & 2.25~s  & 9.25\\
Ramp	& 28950 & 0.520 & -2.16 &  -6.02   & 0.99   & 0.32   & 1.87~s  & 9.77\\
Salaris	& 29200 & 0.525 & -3.45 &  -6.10   & 0.78   & 0.32   & 1.63~s  & 8.81\\
\hline
\multicolumn{9}{c}{}\\
\hline
	& \multicolumn{8}{|c|}{9 periods}	\\
\hline
Model	& \multicolumn{6}{c|}{\teffm [K], \mstarm [\msunm], \Menvm, \Mhem, \Xom, \qfmm} & \sigrmsm &  BIC \\
\hline
Pure C 	& 29650  &  0.530 &   -3.44  &  -6.34   &          &	          & 2.60~s   & 11.3 \\
Ramp	& 29000  &  0.515 &   -3.50  &  -6.90   & 1.00  & 0.31    & 2.04~s   & 11.3 \\
Salaris	& 29300  &  0.510 &   -3.50  &  -7.10   & 0.86  & 0.32    & 2.06~s   & 11.4 \\
\hline
\end{tabular}
\end{table}

We present the best fit parameters for each class of models considered along with a measure of the goodness of each fit in Table \ref{results}. We constrained the 204.6~s and the 281~s modes to be $\ell=1$. We list the periods calculated for the best fit models in the appendix, along with their mode identification. \sigrmsm~is defined as

\begin{equation}
\rm \sigma_{RMS} = \sqrt{\frac{\sum_{1}^{n_{obs}} {(P_{calc}-P_{obs})^2}}{n_{obs}}},
\end{equation}

\noindent where $\rm n_{obs}$ is the number of periods present in the pulsation spectrum. The index labeled "BIC" (Bayes Information Criterion) is a parameter that measures an absolute quality of the fit, by taking into account differing numbers of data points and free parameters \citep[e.g][]{Liddle07}. It penalizes fits involving a greater number of parameters relative to the number of data points. It is given by

\begin{equation}
\rm BIC = n_{obs} \ln(\sigma_{RMS}^2) + n_{par}\ln(n_{obs}),
\end{equation}

\noindent where $n_{par}$ is the number of free parameters in the fit.

\begin{figure}
  \resizebox{\hsize}{!}{\includegraphics{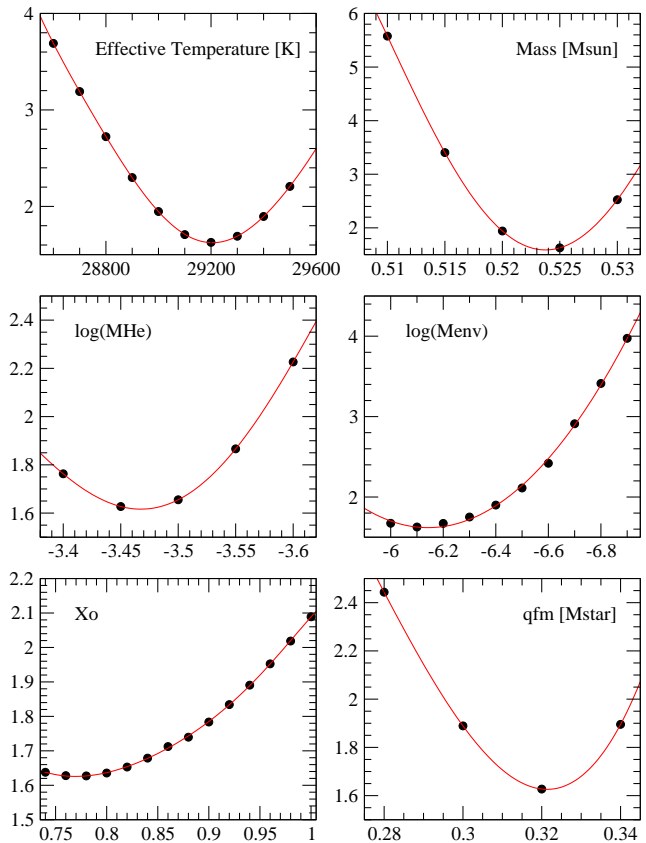}}
  \caption{$\sigma_{RMS}$ as a function of each of the 6 parameters for the Salaris, 8-period best fit model. The vertical axis in each graph is $\sigma_{RMS}$ in seconds. The symbols represent the models calculated, while the lines are polynomial fits to guide the eye.}
  \label{f2}
\end{figure}

From Table \ref{results} we see that in general, the 8 period fits are better than the 9 period fits (even correcting for the fact that it is easier to fit 8 periods than 9) and among the different core profiles, the best fit is the Salaris-like model (with 8 periods). Most models point to a high effective temperature as expected from spectroscopy, and a mass around 0.52 \msunm. The best fits also have a rich to pure oxygen core, and \Menvm $\sim$ -6 to -7. In Fig. \ref{f2} we show how we approach the 8-period Salaris-like best fit model by taking slices through the 6-parameter space at the location of the best fit model (i.e. for the effective temperature plot, we fixed \mstarm~to 0.525 \msunm, \Menvm~to -3.45, \Mhem~to -6.10, \Xom~to 0.78 and \qfmm~to 0.32 while allowing the effective temperature to vary around the best fit value of 29200~K). We show the chemical composition profiles for that model in Fig. \ref{f1}. 

\section{Discussion}
\label{section5}

It is interesting and comforting to note that no matter what the model details are, the best fit parameters are consistent. Parameters that dictate the locations of transition zones, in particular in the core (\qfmm) are especially well determined. The modes are sensitive to the location of chemical transition zones and to the sharpness of the transitions. The sensitivity of the asteroseismic fits to \qfmm~is visible in the last panel of Fig. \ref{f2}. Even with steps of 0.02, the minimum is still barely resolved. The fact that the best fits prefer a high oxygen central abundance (up to pure oxygen) is a sign that the modes are expecting a sharp transition at \qfmm~= 0.31 to 0.34. With our current parameterization of the core composition profiles, the only way to achieve sharper transitions is to increase the oxygen abundance as that allows it to drop more sharply down to zero. It could also explain the preference for the Salaris-like core profiles, as the drop in oxygen at \qfmm~is inherently sharper in these models (see Fig. \ref{f1}). From our results, we can therefore conclude that a sharp composition transition happens at \qfmm~= 0.31 to 0.34, but we have a much weaker constraint on the actual central abundance of oxygen. 

In stellar evolution calculations, the central oxygen abundance for a given mass model is sensitive to the cross section adopted for the $^{12}C(\alpha,\gamma)^{16}O$ reaction (as well as the prescription for convection). The $^{12}C(\alpha,\gamma)^{16}O$ reaction rates measured in the lab also turn out to have fairly large uncertainties. \citet{Metcalfe02} find that these uncertainties propagate into uncertainties in the central oxygen abundance of $\pm 0.1$. There are also uncertainties in the central oxygen abundance associated with the treatment of convection chosen \citep[e.g][]{Metcalfe03b}. All uncertainties considered, the only strong constraint stellar evolution calculations place on the central composition of white dwarfs is that it is roughtly 50/50, with possibly an excess of oxygen over carbon. With the standard reaction rates and treatment for convection, we expect the central abundance for a white dwarf of mass 0.530 \msunm~ to be between $\sim 0.7$ and $\sim 0.9$ \citep{Althaus10,Salaris97}. Our present analysis is approximately consistent with these results, though again, we are more sensitive to the location and shape of the chemical transition zone. 

Diffusion theory predicts that helium diffuses outward, as the heavier elements settle toward the center of the star under the influence of gravity. With time, this leads to a thicker and thicker pure helium layer that sits on top of a mixed core (i.e. \Mhem~gets larger with decreasing effective temperature as the star cools with time). The main question we wanted to answer in this paper, was "Does EC20058-5234 being hotter and younger, have a thinner pure Helium layer than CBS114?". The answer is yes (table \ref{ecvscbs114}). In reaching that conclusion, we must keep in mind that the models used for EC20058-5234 include updated physics. Can we compare the two? We performed 4 parameter fits so we could compare our results with previous asteroseismic fits of EC20058-5234 that were obtained with the same models that were used for the 6 parameter CBS114 fits performed by \citet{Metcalfe05a}. We place these results side by side in Table \ref{ecvscbs114} as well. While the parameters are not identical, we recover the same trends (of particular importance, the helium layer parameters are consistent).

EC20058-5234 also has a lower mass than CBS114 and we need to consider the stellar mass dependence of \Menvm. \citet{Althaus10} find that lower mass models have a thicker helium envelope. Just based on mass, this would mean that EC20058-5234 should have a higher (less negative) \Menvm~than CBS114. That is not the case and we can attribute the thinner helium envelope in EC20058-5234 to the fact that it is younger.

\begin{table}
\label{ecvscbs114}
\caption{EC20058-5234's best fit parameters in the context of testing time-dependent diffusion}
\centering
\begin{tabular}{|l|r|r|r|r|}
\hline
\multicolumn{1}{|c|}{} & \multicolumn{2}{|c|}{Pure carbon EC20058 models} &  \multicolumn{2}{|c|}{Salaris like models} \\
Parameter             & \citet{Metcalfe07} & This work & EC20058 & CBS114 \\
\hline
\teffm [K]              & 28100 & 29600     & 29 200   & 24900 \\
\mstarm [\msunm] & 0.550 & 0.530     & 0.525     & 0.640   \\
\Menvm                 & -3.56  & -3.46      & -3.45      & -2.48   \\
\Mhem                   & -6.42  & -6.26      & -6.10      & -5.94   \\
\Xom                      &           &                & 0.78       & 0.71    \\
\qfmm                       &           &                & 0.32       & 0.38    \\
\hline
\end{tabular}
\end{table}

\section{Conclusions}
\label{section6}

This study reinforces the fact that white dwarf asteroseismology is sensitive to core chemical abundances. In parameterizing composition profiles, questions arise as to how many parameters are needed, how many we can afford to vary (the more we have, the more computationally intensive modeling becomes), and what are the best ways to parameterize the profiles. Because of these difficulties, it is tempting to ignore the importance of properly modeling the core chemical profiles, but the modes carry information about the core structure and we must decipher that information. Stellar evolution calculations are a good starting point and one reason we tried chemical profiles that mimic the ones found by \citet{Salaris97}. More recently, \citet{Althaus10} have performed calculations that evolve stars from the Zero Age Main Sequence to the white dwarf stage, carefully treating massloss and time dependent diffusion of the elements. Pulsating white dwarfs that have over half a dozen modes like EC20058-5234 are good candidates to test these chemical profiles.

In this paper, we used a thorough asteroseismic analysis of EC20058-5234 and compared it with a similar study (varying core parameters) of CBS114 to test diffusion theory in white dwarfs. EC20058-5234 is also astrophysically interesting because it is expected to be hot enough to lose a significant amount of energy through the emission of plasmon neutrinos. While the neutrinos are not detectable by direct methods, they have a significant effect on the rate of cooling of the star. In turn, the rate of cooling can be measured because of its effect on the pulsations. The periods grow longer over time as a result of cooling. The effect is very small ($\sim 10^{-14}$ seconds per second), but detectable over a long enough time. Ideally one can measure an evolutionary cooling rate, as has been done for another pulsating white dwarf, G117-B15A \citep{Kepler05a}. Recently, pulsational data collected over a time period of 13 years for EC20058-5234 was assembled and analyzed to attempt to measure the cooling rate of that star, but the measurement turned out not to be trivial \citep{Dalessio10}. One hypothesis about the strange results is that rotation may play a significant role in the pulsations of the star. This would be consistent with the fact that EC20058-5234 appears to be a fairly fast rotator, with a 2 hr period. Typical (non-magnetic) white dwarfs are known from asteroseismology to rotate with periods $\sim 1$ day \citep[e.g.][]{Kepler03} . The precise mode identification and improved interior models presented here may help shed light on the strange time evolution of EC20058-5234's pulsation periods.

\section*{acknowledgements}
This research was supported by NSF grant AST-0507639 and by the Chemistry, Physics and Astronomy department at Georgia College \& State University. The National Center for Atmospheric Research is sponsored by the U.S.\ National Science Foundation.

\bibliographystyle{mn2e}
\bibliography{index}

\begin{thebibliography}{}

\bibitem[\protect\citeauthoryear{{Althaus}, {C{\'o}rsico}, {Bischoff-Kim},
  {Romero}, {Renedo}, {Garc{\'{\i}}a-Berro} \& {Miller Bertolami}}{{Althaus}
  et~al.}{2010}]{Althaus10}
{Althaus} L.~G.,  {C{\'o}rsico} A.~H.,  {Bischoff-Kim} A.,  {Romero} A.~D.,
  {Renedo} I.,  {Garc{\'{\i}}a-Berro} E.,    {Miller Bertolami} M.~M.,  2010,
  ApJ, 717, 897

\bibitem[\protect\citeauthoryear{{Althaus}, {Serenelli}, {Panei},
  {C{\'o}rsico}, {Garc{\'{\i}}a-Berro} \& {Sc{\'o}ccola}}{{Althaus}
  et~al.}{2005}]{Althaus05}
{Althaus} L.~G.,  {Serenelli} A.~M.,  {Panei} J.~A.,  {C{\'o}rsico} A.~H.,
  {Garc{\'{\i}}a-Berro} E.,    {Sc{\'o}ccola} C.~G.,  2005, A\& A, 435, 631

\bibitem[\protect\citeauthoryear{{Beauchamp}, {Wesemael}, {Bergeron},
  {Fontaine}, {Saffer}, {Liebert} \& {Brassard}}{{Beauchamp}
  et~al.}{1999}]{Beauchamp99}
{Beauchamp} A.,  {Wesemael} F.,  {Bergeron} P.,  {Fontaine} G.,  {Saffer}
  R.~A.,  {Liebert} J.,    {Brassard} P.,  1999, ApJ, 516, 887

\bibitem[\protect\citeauthoryear{{Bischoff-Kim}}{{Bischoff-Kim}}{2008}]{Bischo%
ff-Kim08c}
{Bischoff-Kim} A.,  2008, Communications in Asteroseismology, 154, 16

\bibitem[\protect\citeauthoryear{{Bischoff-Kim}}{{Bischoff-Kim}}{2009}]{Bischo%
ff-Kim09}
{Bischoff-Kim} A.,  2009, in {J.~A.~Guzik \& P.~A.~Bradley} ed., American
  Institute of Physics Conference Series Vol.~1170 of American Institute of
  Physics Conference Series, {Asteroseismological Analysis of Rich Pulsating
  White Dwarfs}.
pp 621--624

\bibitem[\protect\citeauthoryear{{Bradley} \& {Winget}}{{Bradley} \&
  {Winget}}{1994}]{Bradley94}
{Bradley} P.~A.,  {Winget} D.~E.,  1994, ApJ, 430, 850

\bibitem[\protect\citeauthoryear{{Castanheira}, {Kepler}, {Koester} \&
  {Handler}}{{Castanheira} et~al.}{2005}]{Castanheira05}
{Castanheira} B.~G.,  {Kepler} S.~O.,  {Koester} D.,    {Handler} G.,  2005, in
  {Koester} D.,  {Moehler} S.,  eds, 14th European Workshop on White Dwarfs
  Vol.~334 of Astronomical Society of the Pacific Conference Series,
  {Revisiting the DBs Instability Strip Using UV Spectra}.
pp 557--+

\bibitem[\protect\citeauthoryear{{Dalessio}, {Provencal}, {Sullivan} \&
  {Shipman}}{{Dalessio} et~al.}{2010}]{Dalessio10}
{Dalessio} J.,  {Provencal} J.~L.,  {Sullivan} D.~J.,    {Shipman} H.~S.,
  2010, in American Institute of Physics Conference Series American Institute
  of Physics Conference Series, {submitted}

\bibitem[\protect\citeauthoryear{{Dehner} \& {Kawaler}}{{Dehner} \&
  {Kawaler}}{1995}]{Dehner95}
{Dehner} B.~T.,  {Kawaler} S.~D.,  1995, ApJl, 445, L141

\bibitem[\protect\citeauthoryear{{Dreizler} \& {Heber}}{{Dreizler} \&
  {Heber}}{1998}]{Dreizler98}
{Dreizler} S.,  {Heber} U.,  1998, A\&A, 334, 618

\bibitem[\protect\citeauthoryear{{Fontaine} \& {Brassard}}{{Fontaine} \&
  {Brassard}}{2002}]{Fontaine02}
{Fontaine} G.,  {Brassard} P.,  2002, Apj Letters, 581, L33

\bibitem[\protect\citeauthoryear{{Fontaine}, {Graboske} Jr. \& {van
  Horn}}{{Fontaine} et~al.}{1977}]{Fontaine77}
{Fontaine} G.,  {Graboske} Jr. H.~C.,    {van Horn} H.~M.,  1977, ApJs, 35, 293

\bibitem[\protect\citeauthoryear{{Herwig}, {Bl{\"o}cker}, {Langer} \&
  {Driebe}}{{Herwig} et~al.}{1999}]{Herwig99}
{Herwig} F.,  {Bl{\"o}cker} T.,  {Langer} N.,    {Driebe} T.,  1999, A\&A, 349,
  L5

\bibitem[\protect\citeauthoryear{{Iben} Jr., {Kaler}, {Truran} \&
  {Renzini}}{{Iben} et~al.}{1983}]{Iben83}
{Iben} Jr. I.,  {Kaler} J.~B.,  {Truran} J.~W.,    {Renzini} A.,  1983, ApJ,
  264, 605

\bibitem[\protect\citeauthoryear{{Iglesias} \& {Rogers}}{{Iglesias} \&
  {Rogers}}{1996}]{Iglesias96}
{Iglesias} C.~A.,  {Rogers} F.~J.,  1996, ApJ, 464, 943

\bibitem[\protect\citeauthoryear{{Itoh}, {Hayashi}, {Nishikawa} \&
  {Kohyama}}{{Itoh} et~al.}{1996}]{Itoh96}
{Itoh} N.,  {Hayashi} H.,  {Nishikawa} A.,    {Kohyama} Y.,  1996, ApJs, 102,
  411

\bibitem[\protect\citeauthoryear{{Kepler}, {Costa}, {Castanheira}, {Winget},
  {Mullally}, {Nather}, {Kilic}, {von Hippel}, {Mukadam} \&
  {Sullivan}}{{Kepler} et~al.}{2005}]{Kepler05a}
{Kepler} S.~O.,  {Costa} J.~E.~S.,  {Castanheira} B.~G.,  {Winget} D.~E.,
  {Mullally} F.,  {Nather} R.~E.,  {Kilic} M.,  {von Hippel} T.,  {Mukadam}
  A.~S.,    {Sullivan} D.~J.,  2005, ApJ, 634, 1311

\bibitem[\protect\citeauthoryear{{Kepler et al.}}{{Kepler et
  al.}}{2003}]{Kepler03}
{Kepler et al.} 2003, A\& A, 401, 639

\bibitem[\protect\citeauthoryear{{Kilkenny}, {O'Donoghue}, {Crause}, {Hambly}
  \& {MacGillivray}}{{Kilkenny} et~al.}{2009}]{Kilkenny09}
{Kilkenny} D.,  {O'Donoghue} D.,  {Crause} L.~A.,  {Hambly} N.,
  {MacGillivray} H.,  2009, MNRAS, 397, 453

\bibitem[\protect\citeauthoryear{{Koen}, {O'Donoghue}, {Stobie}, {Kilkenny} \&
  {Ashley}}{{Koen} et~al.}{1995}]{Koen95}
{Koen} C.,  {O'Donoghue} D.,  {Stobie} R.~S.,  {Kilkenny} D.,    {Ashley} R.,
  1995, MNRAS, 277, 913

\bibitem[\protect\citeauthoryear{{Lamb} \& {van Horn}}{{Lamb} \& {van
  Horn}}{1975}]{Lamb75}
{Lamb} D.~Q.,  {van Horn} H.~M.,  1975, ApJ, 200, 306

\bibitem[\protect\citeauthoryear{{Lawlor} \& {MacDonald}}{{Lawlor} \&
  {MacDonald}}{2006}]{Lawlor06}
{Lawlor} T.~M.,  {MacDonald} J.,  2006, MNRAS, 371, 263

\bibitem[\protect\citeauthoryear{{Liddle}}{{Liddle}}{2007}]{Liddle07}
{Liddle} A.~R.,  2007, MNRAS, 377, L74

\bibitem[\protect\citeauthoryear{{Metcalfe}}{{Metcalfe}}{2003}]{Metcalfe03b}
{Metcalfe} T.~S.,  2003, ApJl, 587, L43

\bibitem[\protect\citeauthoryear{{Metcalfe}}{{Metcalfe}}{2005}]{Metcalfe05a}
{Metcalfe} T.~S.,  2005, MNRAS, 363, L86

\bibitem[\protect\citeauthoryear{{Metcalfe}}{{Metcalfe}}{2007}]{Metcalfe07}
{Metcalfe} T.~S.,  2007, Communications in Asteroseismology, 150, 227

\bibitem[\protect\citeauthoryear{{Metcalfe} \& {Charbonneau}}{{Metcalfe} \&
  {Charbonneau}}{2003}]{Metcalfe03a}
{Metcalfe} T.~S.,  {Charbonneau} P.,  2003, Journal of Computational Physics,
  185, 176

\bibitem[\protect\citeauthoryear{{Metcalfe}, {Salaris} \& {Winget}}{{Metcalfe}
  et~al.}{2002}]{Metcalfe02}
{Metcalfe} T.~S.,  {Salaris} M.,    {Winget} D.~E.,  2002, ApJ, 573, 803

\bibitem[\protect\citeauthoryear{{Metcalfe}, {Winget} \&
  {Charbonneau}}{{Metcalfe} et~al.}{2001}]{Metcalfe01}
{Metcalfe} T.~S.,  {Winget} D.~E.,    {Charbonneau} P.,  2001, ApJ, 557, 1021

\bibitem[\protect\citeauthoryear{{Miller Bertolami}, {Althaus}, {Serenelli} \&
  {Panei}}{{Miller Bertolami} et~al.}{2006}]{Bertolami06}
{Miller Bertolami} M.~M.,  {Althaus} L.~G.,  {Serenelli} A.~M.,    {Panei}
  J.~A.,  2006, A\&A, 449, 313

\bibitem[\protect\citeauthoryear{{Montgomery}, {Metcalfe} \&
  {Winget}}{{Montgomery} et~al.}{2003}]{Montgomery03}
{Montgomery} M.~H.,  {Metcalfe} T.~S.,    {Winget} D.~E.,  2003, MNRAS, 344,
  657

\bibitem[\protect\citeauthoryear{{Nather}}{{Nather}}{1989}]{Nather89}
{Nather} R.~E.,  1989, in {Wegner} G.,  ed., IAU Colloq. 114: White Dwarfs
  Vol.~328 of Lecture Notes in Physics, Berlin Springer Verlag, {The whole
  earth telescope}.
pp 109--114

\bibitem[\protect\citeauthoryear{{Nitta}, {Kleinman}, {Krzenski}, {Kepler},
  {Metcalfe}, {Mukadam}, {Mullally}, {Nather}, {Sullivan}, {Thompson} \&
  {Winget}}{{Nitta} et~al.}{2009}]{Nitta09}
{Nitta} A.,  {Kleinman} S.~J.,  {Krzenski} J.,  {Kepler} S.~O.,  {Metcalfe}
  T.~S.,  {Mukadam} A.~S.,  {Mullally} F.,  {Nather} R.~E.,  {Sullivan} D.,
  {Thompson} S.~E.,    {Winget} D.~E.,  2009, Journal of Physics Conference
  Series, 172, 012073

\bibitem[\protect\citeauthoryear{{Salaris}, {Dominguez}, {Garcia-Berro},
  {Hernanz}, {Isern} \& {Mochkovitch}}{{Salaris} et~al.}{1997}]{Salaris97}
{Salaris} M.,  {Dominguez} I.,  {Garcia-Berro} E.,  {Hernanz} M.,  {Isern} J.,
    {Mochkovitch} R.,  1997, ApJ, 486, 413

\bibitem[\protect\citeauthoryear{{Saumon}, {Chabrier} \& {van Horn}}{{Saumon}
  et~al.}{1995}]{Saumon95}
{Saumon} D.,  {Chabrier} G.,    {van Horn} H.~M.,  1995, ApJs, 99, 713

\bibitem[\protect\citeauthoryear{{Sullivan}, {Metcalfe} \&
  {O'Donoghue}}{{Sullivan} et~al.}{2008}]{Sullivan08}
{Sullivan} D.~J.,  {Metcalfe} T.~S.,    {O'Donoghue} D. e.~a.,  2008, MNRAS,
  387, 137

\bibitem[\protect\citeauthoryear{{Wood}}{{Wood}}{1990}]{Wood90}
{Wood} M.~A.,  1990, PhD thesis, AA(Texas Univ., Austin.)

\bibitem[\protect\citeauthoryear{{Yeates}}{{Yeates}}{2006}]{Yeates06}
{Yeates} C.~M.,  2006, PhD thesis, The University of North Carolina at Chapel
  Hill

\end{thebibliography}

\appendix
\section{Model periods and mode identification for EC20058-5234}
We list here the periods calculated for the best fit models presented in the main text, along with their mode identification. The information contained in this appendix shows in more detail the quality of the fits and may prove useful for further asteroseismic studies of EC20058-5234. We stress again that the 204.6~s and 281.0~s modes were constrained to be $\ell=1$ while the others were allowed to be either $\ell=1$ or $\ell=2$.

For all best fit models, the 350.6 second mode has a good fit either as $\ell=1$ (k = 5) or $\ell=2$ (k = 13). In all cases, the two periods are less than 2 seconds apart and in most cases, less than a second apart. In other words, the two identifications are not distinguishable from one another. We chose to list both possibilities in tables \ref{csal8} and \ref{cramp9} (so the 350.6 second mode appears twice in each table even though it is a single mode).

\begin{table}
\caption{
\label{csal8}
Periods and mode identification for the 8-period fits}
\centering
\begin{tabular}{|c|c|c|c|c|c|}
\hline
Observed Period [s]         & \multicolumn{3}{|c|}{Model Period [s]} & $\ell$ & k   \\
                                         & Pure C          & Ramp         & Salaris       &           &      \\
\hline
204.6                                & 204.7            & 204.5         & 203.6                       & 1       & 3   \\
281.0                                & 280.1            & 279.4         & 281.7                       & 1       & 5   \\
350.6                                & 350.8            & 349.1         & 350.2                       & 1       & 7   \\
\hline
195.0                                & 191.9            & 192.6         & 192.3                       & 2       & 6   \\
256.9                                & 256.9            & 257.0         & 259.6                       & 2       & 9   \\
333.5                                & 330.5            & 330.5         & 331.4                       & 2       & 12 \\
350.6                                & 349.0            & 350.5         & 351.2                       & 2       & 13 \\
525.4                                & 524.6            & 522.4         & 525.5                       & 2       & 21 \\
539.8                                & 541.4            & 541.1         & 540.6                       & 2       & 22 \\
\hline
\end{tabular}
\end{table}

\begin{table}
\caption{
\label{cramp9}
Periods and mode identification for 9 period fits}
\centering
\begin{tabular}{|c|c|c|c|c|c|}
\hline
Observed Period [s]         & \multicolumn{3}{|c|}{Model Period [s]} & $\ell$ & k   \\
                                         & Pure C          & Ramp         & Salaris       &           &      \\
\hline
204.6                                & 204.4            & 204.9         & 205.3        & 1       & 3   \\
281.0                                & 279.9            & 282.5         & 282.4        & 1       & 5   \\
350.6                                & 350.1            & 349.3         & 350.0        & 1       & 7   \\
\hline
195.0                                & 191.4            & 192.8         & 193.0        & 2       & 6   \\
256.9                                & 260.2            & 258.9         & 257.6        & 2       & 9   \\
274.7                                & 278.9            & 276.8         & 277.5        & 2       & 10 \\
333.5                                & 329.7            & 330.0         & 329.0        & 2       & 12 \\
350.6                                & 348.3            & 349.1          & 350.3       & 2        & 13 \\
525.4                                & 523.9            & 523.8         & 525.3        & 2       & 21 \\
539.8                                & 541.0            & 542.2         & 541.5        & 2       & 22 \\
\hline
\end{tabular}
\end{table}

\label{lastpage}

\end{document}